\documentclass[a4paper,12pt]{article}
\def\letter{0}\def\pr{0}
\pdfoutput=1 
\usepackage{simplewick}
\usepackage{graphicx}
\usepackage{epstopdf}
\usepackage{ifthen}
\usepackage{csquotes}
\usepackage[numbers,sort&compress]{natbib}
\usepackage{tikz}
\usetikzlibrary{arrows.meta} 

\usepackage{amsmath}
\usepackage[psamsfonts]{amssymb}
\usepackage{euscript}
\usepackage{caption}
\usepackage{latexsym}
\usepackage[arrow,matrix,curve]{xy}

\usepackage[hypertexnames=false]{hyperref}
\hypersetup
{
    colorlinks=true,
    linkcolor=blue,
    filecolor=magenta,   
    citecolor=black,   
    urlcolor=cyan,
}

\def\,{\hspace{-.1cm}}
\def\hsp{,\hspace{.7cm}}

\def\fc#1#2 {\frac{n}{q}#1\frac{n}{q}#2}

\newcommand{\vac}{\ensuremath{|0\rangle}}

\renewcommand{\cos}{\textrm{cos}}
\renewcommand{\sin}{\textrm{sin}}

\renewcommand{\sinh}{\textrm{sinh}}
\renewcommand{\cosh}{\textrm{cosh}}
\renewcommand{\tanh}{\textrm{tanh}}
\newcommand{\sech}{\textrm{sech}}
\newcommand{\csch}{\textrm{csch}}

\def\exp#1{\hbox{\rm exp}\left[#1\right]}

\renewcommand{\theequation}{\arabic{section}.\arabic{equation}}
\renewcommand{\(}{\begin{equation}}
\renewcommand{\)}{end{equation} \vspace{-.05in}\linebreak}

\newcounter{saveeqn}
\newcounter{savealpheqn}

\newcommand{\alpheqn}{\setcounter{saveeqn}{\value{equation}}%
  \stepcounter{saveeqn}\setcounter{equation}{0}%
  \renewcommand{\theequation}{\mbox{\arabic{section}.\arabic{saveeqn}
\alph{equation}}}
  \renewcommand{\)}{\end{equation}}}
\def\part#1{\frac{\partial}{\partial{#1}}}%
\def\group#1{\refstepcounter{equation}\setcounter{saveeqn}
 {\value{equation}}%
  \label{#1}\setcounter{equation}{0}%
\renewcommand{\theequation}{\mbox{\arabic{section}.\arabic{saveeqn}
\alph{equation}}}
  \renewcommand{\)}{\end{equation}}}
\newcommand{\reseteqn}{\setcounter{equation}{\value{saveeqn}}%
  \renewcommand{\theequation}{\arabic{section}.\arabic{equation}}%
  \renewcommand{\)}{\end{equation}}}

\newcommand{\aalpheqn}{\setcounter{saveeqn}{\value{equation}}%
  \stepcounter{saveeqn}\setcounter{equation}{0}%
  \renewcommand{\theequation}{\mbox{
        \Alph{subsection}.\arabic{saveeqn}\alph{equation}}}
   \renewcommand{\)}{\end{equation}}}
\newcommand{\areseteqn}{\setcounter{equation}{\value{saveeqn}}%
  \renewcommand{\theequation}{\Alph{subsection}.\arabic{equation}}%
  \renewcommand{\)}{\end{equation}}}

\renewcommand{\thefootnote}{\alph{footnote}}
\renewcommand{\(}{\begin{equation}}
\renewcommand{\)}{\end{equation}}
\newcommand{\ba}{\begin{eqnarray}}
\newcommand{\ea}{\end{eqnarray}}
\renewcommand{\a}{\alpha}
\renewcommand{\b}{\beta}

\renewcommand{\sl}{{\sqrt{\lambda}}}

\newcommand{\cbp}{\mathop{\vtop{\ialign{##\crcr
   $\hfil\displaystyle{}\hfil$\crcr\noalign{\kern-13pt\nointerlineskip}
   \BIG{)}\hskip0pt\crcr\noalign{\kern3pt}}}}}
\newcommand{\pa}{\mathop{\vtop{\ialign{##\crcr

$\hfil\displaystyle{\oplus}\hfil$\crcr\noalign{\kern+1pt\nointerlineskip
}
   \hspace{.08in}$^{\alpha=0}$\hskip6pt\crcr\noalign{\kern3pt}}}}}
\renewcommand{\hsp}{,\hspace{.3in}}
\newcommand{\p}{^\prime}

\def\D{\ensuremath{{\cal D}}}

\catcode`\@=11
\def\vereq#1#2{\lower3pt\vbox{\baselineskip1.5pt \lineskip1.5pt
\ialign{$\m@th#1\hfill##\hfil$\crcr#2\crcr\sim\crcr}}}
\catcode`\@=12

\renewcommand{\(}{\begin{equation}}
\renewcommand{\)}{\end{equation}}

\def\k#1{\k_{#1}}

\def\hc{{\bf{H}}}
\def\hq{{\bf {\hat{H}}}}
\def\xb{\left(\frac{x}{\sqrt\beta}\right)}
\def\pb{\left(\frac{\phi(x)}{\sqrt\beta}\right)}

\def\g{\mathfrak g}

\def\red#1{\textcolor{red}{hengyuan: #1}}

\def\gre#1{\textcolor{magenta}{jarah: #1}}
\usepackage[dvipsnames]{xcolor}

\newcommand{\beas}{\begin{eqnarray*}}
\newcommand{\eeas}{\end{eqnarray*}}

\newcommand{\bquo}{\begin{quote}}
\newcommand{\enqu}{\end{quote}}

\def\lim#1{\stackrel{\rm{lim}}{{}_{#1}}}

\def\ch{{\mathcal{H}}}

\def\x#1{\left(\frac{mx}{#1\hbar}\right)}

\newcommand{\beq}{\begin{equation}}
\newcommand{\eeq}{\end{equation}}
\newcommand{\bea}{\begin{eqnarray}}
\newcommand{\eea}{\end{eqnarray}}

\newif\ifdtup

\catcode`\@=11

\ifthenelse{\equal{\letter}{0}}
{ 

\@addtoreset{equation}{section}
\def\theequation{\arabic{section}.\arabic{equation}}

\def\@normalsize{\@setsize\normalsize{15pt}\xiipt\@xiipt
\abovedisplayskip 14pt plus3pt minus3pt%
\belowdisplayskip \abovedisplayskip
\abovedisplayshortskip \z@ plus3pt%
\belowdisplayshortskip 7pt plus3.5pt minus0pt}

\def\small{\@setsize\small{13.6pt}\xipt\@xipt
\abovedisplayskip 13pt plus3pt minus3pt%
\belowdisplayskip \abovedisplayskip
\abovedisplayshortskip \z@ plus3pt%
\belowdisplayshortskip 7pt plus3.5pt minus0pt
\def\@listi{\parsep 4.5pt plus 2pt minus 1pt
      \itemsep \parsep
      \topsep 9pt plus 3pt minus 3pt}}

\relax

\def\section{\@startsection{section}{1}{\z@}{3.5ex plus 1ex minus  .2ex}{2.3ex plus .2ex}{\large\bf}}

\def\thesection{\arabic{section}}
\def\thesubsection{\arabic{section}.\arabic{subsection}}

\def\appendix{\setcounter{section}{0}
 \def\thesection{Appendix \Alph{section}}
 \def\thesubsection{\Alph{section}.\arabic{subsection}}
 \def\theequation{\Alph{section}.\arabic{equation}}}
\renewcommand{\theequation}{\arabic{section}.\arabic{equation}}
}
{
\renewcommand{\theequation}{\arabic{equation}}
} 

\begin{document}
\def\thefootnote{\fnsymbol{footnote}}
\def\thetitle
{Quantum Lifts of Noninteger Power Law Field Theories}
\def\autone{Hengyuan Guo}
\def\auttwo{Jarah Evslin}
\def\autthree{Stefano Bolognesi}


\def\affa{ School of Physics and Astronomy, Sun Yat-sen University, Zhuhai 519082, China}
\def\affaa{Department of Physics, ``Enrico Fermi'', University of Pisa; INFN, Sezione di Pisa, \\ Largo Pontecorvo, 3, 56127, Pisa, Italy}
\def\affaaa{Lanzhou Center for Theoretical Physics, \\ Key Laboratory of Theoretical Physics of Gansu Province, \linebreak Key Laboratory of Quantum Theory and Applications of MoE,  Lanzhou University, \\ Lanzhou, Gansu 730000, China}
\def\affb{Institute of Modern Physics, NanChangLu 509, Lanzhou 730000, China}
\def\affc{University of the Chinese Academy of Sciences, YuQuanLu 19A, Beijing 100049, China}



\ifthenelse{\equal{\pr}{1}}
{
\title{\thetitle}
\author{\auttwo}
\author{\autone}
\author{\autthree}
\affiliation {\affb}
\affiliation {\affc}
\affiliation {\affa}
\affiliation {\affaa}
\affiliation {\affaaa}
}

\begin{center}
{\large {\bf \thetitle}}

\bigskip

\large \noindent  
\auttwo$^{1,2}$ 
\footnote{jarah@impcas.ac.cn}, 
\autone$^{3,4,5}$  
\footnote{guohy57@mail.sysu.edu.cn}  
and  
\autthree$^{4}$
\footnote{stefano.bolognesi@unipi.it}

\vskip.7cm  
{\small 
1) \affb\\
2) \affc\\
3) \affa\\
4) \affaa\\
5) \affaaa\\
}
\end{center}

\begin{abstract}
\noindent
Field theories whose potentials have noninteger power laws $\alpha$ have found many applications, but are often claimed to have no lift to quantum field theory except as effective models.  We define quantum lifts by expanding the classical potential in Hermite polynomials and then normal ordering at a mass scale shifted by a parameter $\beta$.  We find that when $\alpha>2$, for sufficiently large $\beta$, the vacuum state can be perturbatively expanded in usual Fock states.  We apply this to the following problem.  The $\sigma=4$ P\"oschl-Teller model has a $\phi^{5/2}$ potential.  As the third derivative of the potential diverges in each vacuum, one expects the three point interactions to diverge in the vacuum.  The model's kink has three shape modes and the least bound mode extends so far into the vacuum that its probability of being excited by radiation apparently diverges.  We show that a deformation $\beta$ of order the meson mass or larger is sufficient to tame this divergence, although it nonetheless results in an excitation probability which is enhanced by a $\beta$-dependent fractional power of the inverse coupling.

\end{abstract}

%
\setcounter{footnote}{0}
\renewcommand{\thefootnote}{\arabic{footnote}}

\ifthenelse{\equal{\pr}{1}}
{
\maketitle
}{}

\section{Introduction}

\subsection{Power Law Potentials}

In the twentieth century it was often claimed that fundamental theories should have analytic potentials \cite{sg2}.  Nonetheless, nonanalytic models have been indispensable from the days of Einstein-Hilbert and Born-Infeld actions to those of the Nambu-Goto action to the Signum-Gordon model \cite{signum} of today.  Indeed, models with nonanalytic potentials have many applications \cite{frac1,frac2,frac3,frac4,frac5} and admit new solutions, such as the compacton \cite{comp02}. As a result, their popularity has only increased in the past year \cite{conf25,na4}.  This situation seems set to continue.  Indeed, if the next generation of cosmic microwave background probes do not discover primordial tensor perturbations, then the only power law inflationary models left will have powers less than unity \cite{ali22}.  In some applications, such as in condensed matter physics, these models clearly are low energy effective theories.  Nonetheless, the increasing popularity of such models warrants an investigation of just how they are to be treated in the deep quantum regime.

In the present note we consider a classical scalar field theory in 1+1 dimensions with a mass term and a potential of the form $\lambda|x^\alpha|$ with $\alpha$ not an integer.  For any integer $n>\alpha$, the $n$th derivative of the potential diverges and so one may worry that the $n$-point function itself is ill-defined.  We claim that, as usual, there exist many distinct quantum lifts.  We describe a family of lifts indexed by a real parameter $\beta$ with units of action.  This $\beta$ is defined to be a shift in the Wick contraction of the scalar field with itself, but can equivalently be defined to be a shift in the mass scale of the normal ordering \cite{no}.  We find that, when $\alpha$ and $\beta/\hbar$ are large enough, the $n$-point functions are described by the usual $:\phi^n(x):$ interactions with finite coefficients that depend on the choice of $\beta$.  Thus these quantum lifts of the power law models appear to be sensible, however they are not unique.

\subsection{An Application}

We are motivated by the following problem.  In Refs.~\cite{sg2,tf} the authors introduced the P\"oschl-Teller models, which, at level $\sigma$ enjoy kink solutions whose perturbations feel level $\sigma$ P\"oschl-Teller potentials.  Expanding about a minimum of the potential, the leading interaction term in these models is proportional to $|\phi^{2+2/\sigma}(x)|$, whose third derivative is not defined when $\sigma>2$, suggesting that their three-point couplings diverge.  As a result, both references claimed that when $\sigma>2$ these models are unphysical.  The second specified that even the one-loop mass is unlikely to exist, although it was soon shown that this prediction was incorrect \cite{boya}.  

But what about processes that make explicit use of the three-point coupling?  The divergence of the third derivative happens in the vacuum, and so near a kink the third derivative is finite and one expects no problems.  Therefore in Ref.~\cite{hengyuanstokes} we calculated, in the case $\sigma=3$, the amplitude of the process in which incoming radiation excites a mode which is bound to a kink.  The interaction does appear to grow with the distance from the kink, suggesting that incoming radiation interacts more strongly with the kink the further it is.  However, this growth is compensated by the fact that the shape modes are tightly bound, and so are exponentially suppressed far from the kink.  Therefore no divergence arose.

What about the case $\sigma=4$?  This model has three shape modes and one is quite loosely bound.  A calculation similar to that done for $\sigma=3$ shows that this loosely bound mode extends so far into the bulk that its exponential suppression at long distances is not sufficient to compensate for the increased third derivative, and so the divergence remains.

We will show that when $\beta$ is sufficiently large it cuts off this long distance divergence in the amplitude for the shape mode to be excited by incoming radiation.  This can be done with $\beta$ still sufficiently small that the semiclassical limit yields the classical P\"oschl-Teller model.  The excitation amplitude becomes finite, however it is nonetheless dominated by interactions of the meson with the kink when their separation is relatively large, where the shape mode itself is exponentially suppressed and the coupling is exponentially enhanced.  This contribution in fact dominates over the usual interaction arising from the cubic vertex, implying that the excitation probability is not of order $O(\lambda)$ but in fact of a lower, nonintegral power of $\lambda$ which depends on the choice of $\beta$.  Thus the usually perturbative expansion in powers of $\lambda$ is replaced by an expansion in noninteger powers.

\section{Quantum Mechanics}

\subsection{The Construction}

Consider a classical scalar $x$ in 0+1 dimensions.  Classically, it is described by the Hamiltonian
\beq
\hc=\hc_0+\hc_I\hsp \hc_0=\frac{p^2+\omega^2 x^2}{2}\hsp \hc_I=\lambda |x^\alpha|.
\eeq
To quantize, let us impose the canonical quantization relations
\beq
[x,p]=i\hbar.
\eeq
As usual, one may define the linear combination
\beq
a=\sqrt{\frac{\omega}{2\hbar}}x+\frac{ip}{\sqrt{2\hbar\omega}}.
\eeq
One lift of the free Hamiltonian to a quantum operator is
\beq
\hq_0=\hbar\omega a^\dag a=\frac{p^2+\omega^2x^2}{2}-\frac{\hbar\omega}{2}.
\eeq
Note that, setting $\hbar=0$, this is indeed equal to the free part of the classical Hamiltonian $\hc_0$.

Let us introduce
\beq
\beta=\frac{\hbar}{\omega} \label{bdef}
\eeq
which is the variance of $x$ in the ground state $\hq_0$
\beq
\vac_0=\frac{1}{(\pi\beta)^{1/4}}\int dx e^{-x^2/(2\beta)}|x\rangle\hsp \langle x|y\rangle=\delta(x-y).
\eeq
Then the other eigenstates of $\hq_0$ are
\beq
|n\rangle_0=\frac{a^{\dagger n}\vac_0}{\sqrt{n!}}.
\eeq
The numerator can be expressed in terms of Hermite polynomials $H_n$ as
\beq
a^{\dag n}\vac_0=2^{-n/2}H_n\xb \vac_0.
\eeq
Defining normal-ordering, which places all $a^\dag$ to the left of $a$, one can in turn express the Hermite polynomials as
\beq
:\xb^n:=2^{-n}H_n\xb.
\eeq
This can be shown recursively, using the identity
\beq
x^2-:x^2:=\frac{\hbar}{2\omega}[a,a^\dag]=\frac{\beta}{2}.
\eeq

Can we lift $\hc_I$ to a quantum operator?  To do this, let us first expand $\hc_I$ in Hermite polynomials using
\beq
|x|^\alpha=\sum_{n=0}^\infty c_{2n,\alpha} H_{2n}(x)\hsp
c_{2n,\alpha}=\frac{1}{\sqrt{\pi}(2n)!}\left(\frac{\alpha}{2}\right)_n\Gamma\left(\frac{\alpha+1}{2}\right) \label{oexp}
\eeq
as is shown in \ref{happ}.  Here $()_n$ denotes the falling factorial.  This series expansion converges pointwise in $x$. 
\begin{figure}
    \centering    \includegraphics[width=0.7\linewidth]{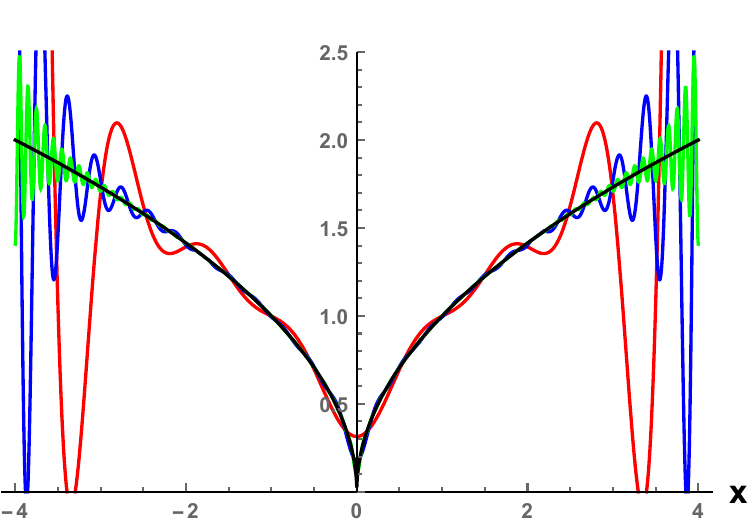} 
    \caption{The convergence of the expansion of Eq.~(\ref{oexp}) for $|x|^{1/2}$ (black) including the first 10 (red), 100 (blue) and 1000 (green) terms}
    \label{sqrt}
\end{figure}

While the classical $\hc_0$ was related to the quantum $\hq_0$ by a shift $\hbar\omega/2$, in the case of $\hc_I$ we will not use any shift at all.  As the classical interaction is
\beq
\hc_I=\lambda |x^\alpha|=\lambda \beta^{\alpha/2} \left|\xb^\alpha\right|
\eeq
we will simply define a series of quantum Hamiltonians
\bea
\hq_{I,N}&=&\lambda\beta^{\alpha/2} \sum_{n=0}^N c_{2n,\alpha} H_{2n}\xb=\lambda\beta^{\alpha/2} \sum_{n=0}^N 2^{2n} c_{2n,\alpha} :\xb^{2n}:\\
&=&\frac{\lambda \beta^{\alpha/2} \Gamma\left(\frac{\alpha+1}{2}\right)}{\sqrt{\pi}}\left[ 
1+\frac{\alpha}{\beta}:x^2:+\frac{\alpha(\alpha-2)}{6\beta^2}:x^4:+\frac{\alpha(\alpha-2)(\alpha-4)}{90\beta^3}:x^6:+O\left(\frac{x^8}{\beta^4}\right)
\right]\nonumber
\eea
indexed by $N$.  The action of $\hq_{I,N}$ on a state converges pointwise in $x$.  
Intuitively, as the operator on the last line is normal-ordered, its action on each state with finite occupation number is not so large and, since $c_{2n,\alpha}$ shrinks exponentially in $n$, this action on any state whose energy is of order $O(\hbar\omega)$ converges in $N$.  

In addition to the convergence provided by the shrinking $c_{2n,\alpha}$, note that each time $n$ increases by one, the last expression on the first line has an additional power of $x^2/\beta$.  In the case of superpositions of excitations up to some fixed excitation number, the wave functions have support at $x\sim O(\sqrt{\hbar/\omega})$ and so this factor is $\hbar/(\beta \omega)$.  It is equal to unity.  However, if somehow we could instead take the limit $\hbar/(\beta\omega)\rightarrow 0$, then this series would define a perturbative expansion in $\hbar/(\omega \beta)$.  In quantum field theory, we will do this by modifying the Hamiltonian.

\subsection{Example}
As a simple example of this nice behavior, let us provide the leading correction $\vac_1$ to the ground state $\vac_0$.  As the energy of the ground state is zero, this correction satisfies the Schr\"odinger equation
\beq
\hq_0\vac_1=-\hq_{I,N}\vac_0. \label{seq}
\eeq

The right hand side is
\bea
-\hq_{I,N}\vac_0&=&
-\lambda\beta^{\alpha/2} \sum_{n=0}^N 2^{2n} c_{2n,\alpha} :\xb^{2n}:\vac_0\\
&=&
-\lambda\beta^{\alpha/2} \sum_{n=0}^N 2^{n}c_{2n,\alpha}a^{\dagger 2n}\vac_0\nonumber\\
&=&
-\lambda\beta^{\alpha/2} \sum_{n=0}^N 2^{n} c_{2n,\alpha}\sqrt{(2n)!}|2n\rangle_0.\nonumber
\eea

Inserting
\beq
\hq_0|n\rangle_0=n\hbar\omega|n\rangle_0
\eeq
into the left hand side of Eq.~(\ref{seq}), we conclude that the leading perturbative correction to the ground state is
\beq
\vac_1=-\lambda\beta^{\alpha/2} \sum_{n=0}^N \frac{2^{n-1} c_{2n,\alpha}\sqrt{(2n)!}}{n\hbar \omega}|2n\rangle_0.
\eeq
In particular, in passing from $\hq_{I,N-1}$ to $\hq_{I,N}$ the correction $\vac_1$ changes by
\beq
\Delta \vac_1=-\lambda\beta^{\alpha/2} \frac{2^{N-1} c_{2N,\alpha}\sqrt{(2N)!}}{N\hbar \omega}|2N\rangle_0
\eeq
which has norm
\beq
|\Delta\vac_1|^2=\lambda^2\hbar^{\alpha-2} \frac{2^{2N-2} c^2_{2N,\alpha}(2N)!}{N^2 \omega^{\alpha+2}}=\frac{\lambda^2}{\pi\omega^{\alpha+2}}\Gamma^2\left(\frac{\alpha+1}{2}\right)\hbar^{\alpha-2} \frac{2^{2N-2} }{N^2(2N)! }
\left[\left(\frac{\alpha}{2}\right)_N\right]^2.
\eeq
One sees that for positive $\alpha$ this series indeed converges, and for $\alpha>2$, this correction is subdominant to $\vac_0$ in the semiclassical expansion in $\hbar$.  Thus, at least for $|x|$ not too large, the states seem to have a healthy semiclassical expansion when $\alpha>2$.

\section{Quantum Field Theory}

\subsection{The Construction}

Let us consider a classical theory of a scalar field in 1+1 dimensions, defined by the Hamiltonian density
\beq
\ch(x)=\ch_0(x)+\ch_I(x)\hsp \ch_0(x)=\frac{\pi^2(x)+\left(\partial_x\phi(x)\right)^2+\frac{m^2}{\hbar^2}\phi^2(x)}{2}\hsp
\ch_I(x)=\lambda|\phi^\alpha(x)|. \label{classh}
\eeq
Note that $m/\hbar$ is a frequency which tends to a finite value in the semiclassical limit $m,\hbar\rightarrow 0$ of the corresponding quantum theory, and that limiting value appears in the Hamiltonian of the classical theory (\ref{classh}).

To quantize, we will impose the canonical commutation relation
\beq
[\phi(x),\pi(y)]=i\hbar \delta(x-y).
\eeq
It will be convenient to introduce the linear combinations
\beq
a_p=\int dx e^{ipx/\hbar}\left[ 
\sqrt\frac{\omega_p}{2\hbar}\phi(x)+\frac{i}{\sqrt{2\hbar\omega_p}}\pi(x)\right]\hsp \omega_p=\frac{1}{\hbar}\sqrt{m^2+p^2}
\eeq
so that
\beq
[a_p,a^\dag_q]= 2\pi\delta(p-q). 
\eeq
To define a quantum field theory, we should choose a quantum Hamiltonian.  For the free part, we make the conventional choice
\beq
\hat\ch_0(x)=:\ch_0(x):
\eeq
where the normal ordering :: places all $a^\dag_p$ to the left of $a_q$.  We will use the notation $\vac_0$ for the vacuum of the free Hamiltonian.

For the interaction term, for some $\beta$ to be determined, we will first expand the classical interaction term
\beq
\ch_I(x)=\lambda|\phi^\alpha(x)|=\lambda \beta^{\alpha/2} \left|\pb^\alpha\right|=\lambda\beta^{\alpha/2} \sum_{n=0}^\infty c_{2n,\alpha} H_{2n}\pb\label{chc}.
\eeq
We would like to use a $\beta$ normal ordering prescription $::_\beta$ which satisfies the identity
\beq
:H_{n}\pb:_\beta=2^{n} :\pb^{n}:\label{sogno}
\eeq
where on the right hand side we have used the usual normal ordering prescription.  

How is $::_\beta$ to be defined in general? First let us define a $\beta$-contraction
\beq
\contraction{}{\phi}{(x)}{\phi}
  \phi(x)\phi(x)=\frac{\beta}{2}.
\eeq
Then, for a bilinear in the fields, the relationship between the $\beta$ normal ordering and the usual normal ordering is defined to be
\beq
:\phi^2(x):_\beta=:\phi^2(x):+\contraction{}{\phi}{(x)}{\phi}
  \phi(x)\phi(x)=:\phi^2(x):+\frac{\beta}{2}. \label{andef}
\eeq
Our definition of the normal ordering $::_\beta$ for a general polynomial will be the linear operation such that each monomial, $:\phi^n(x):_\beta$ is defined by summing, as in Wick's theorem, all combinations of Wick contractions to $:\phi^n:$, replacing each contracted pair of $\phi(x)$ with $\beta/2$. For example
\bea
:\phi^3(x):_\beta&=&:\phi^3(x):+\contraction{}{\phi}{(x)}{\phi}
  \phi(x)\phi(x):\phi(x):
  +\contraction{}{\phi}{(x):\phi(x):}{\phi}
  \phi(x):\phi(x):\phi(x)
  +:\phi(x):\contraction{}{\phi}{(x)}{\phi}
  \phi(x)\phi(x)\nonumber\\
  &=&:\phi^3(x):+\frac{3\beta}{2}\phi(x).
\eea
In other words, the usual divergent contraction $\phi^2-:\phi^2:$ is shifted by a finite amount $\beta/2$.  While $\beta$ is still arbitrary, the condition that normal ordering does not affect the classical limit implies that $\beta\rightarrow 0$ when $\hbar\rightarrow 0$.  For example, one may choose $\beta=\hbar$.

Note that, following Ref.~\cite{colemansg}, this shift in the normal ordering is equivalent to a shift
\beq
\Delta m=\pi m \frac{\beta}{\hbar} +O\left(m\frac{\beta^2}{\hbar^2}\right)
\eeq
of the mass scale used to define the normal ordering.  In other words, the normal ordering $::_\beta$ is equivalent to the usual normal ordering $::$, but performed at a mass scale not equal to $m$.

Now we are ready to define the quantum interaction Hamiltonian
\beq
\hat\ch_{I,N}(x)=\lambda\beta^{\alpha/2} \sum_{n=0}^N c_{2n,\alpha} :H_{2n}\pb:_\beta.
\eeq
As we have demanded that $\beta\rightarrow 0$ in the semiclassical limit $\hbar\rightarrow 0$, for any such $\beta$ this is indeed a quantum lift of our classical interaction.

Inserting Eq.~(\ref{sogno}) we find
\bea
\hat\ch_{I,N}(x)&=&\lambda\beta^{\alpha/2} \sum_{n=0}^N 2^{2n} c_{2n,\alpha}:\pb^{2n}:\\
&=&\frac{\lambda \beta^{\alpha/2} \Gamma\left(\frac{\alpha+1}{2}\right)}{\sqrt{\pi}}\left[ 
1+\frac{\alpha}{\beta}:\phi^2(x):+\frac{\alpha(\alpha-2)}{6\beta^2}:\phi^4(x):+O\left(\frac{\phi^6(x)}{\beta^3}\right)
\right]\nonumber
\eea
 Note that each time $n$ increases by one, the expression on the right hand side of the first line has an additional power of $\phi^2/\beta$.  In the multiparticle Fock space, the state has support at $\phi(x)\sim O(\sqrt{\hbar})$ and so this factor is $\hbar/\beta$.  Now let us take the limit $\hbar/\beta\rightarrow 0$.  In this case, this series defines a perturbative expansion in $\hbar/\beta$.

\subsection{Example}

We have defined $\vac_0$ to be the lowest energy eigenstate of the free Hamiltonian.  One may attempt to use perturbation theory to calculate the vacuum of the full, interacting Hamiltonian.  Let us assume for now that $\vac_0$ is the leading term, and let $\vac_1$ be the subleading term.  These are related by the Schr\"odinger equation
\beq
\hat{H}_0\vac_1=-\hat{H}_{I,N}\vac_0\hsp \hat{H}_0=\int dx \ch_0(x)\hsp \hat{H}_{I,N}=\int dx \ch_{I,N}(x). 
\eeq

Let us define the shorthand
\beq
A_p=\sqrt{2\omega_p}a_p\hsp A^\ddag_p=\frac{a^\dag_p}{\sqrt{2\omega_p}}\hsp |p_1\cdots p_n\rangle_0=A^\ddag_{p_1}\cdots A^\ddag_{p_n}\vac_0.
\eeq
Note that
\beq
{}_0\langle 0\vac_0=0\hsp {}_0\langle p|q\rangle_0=\frac{2\pi\delta(p-q)}{2\omega_q}.
\eeq

Using
\beq
\int dx :\phi^n(x):\vac_0=\hbar^{n/2}\int\frac{d^{n-1}p}{(2\pi)^{n-1}}|p_1,\cdots,p_{n-1},-\sum_{i=1}^{n-1}p_i\rangle_0
\eeq

we evaluate the right hand side of the Schr\"odinger equation
\beq
-\hat{H}_{I,N}\vac_0=-\lambda \sum_{n=0}^N  2^{2n} \beta^{\frac{\alpha}{2}-n}\hbar^n c_{2n,\alpha}\int\frac{d^{2n-1}p}{(2\pi)^{2n}}|p_1,\cdots,p_{2n-1},-\sum_{i=1}^{2n-1}p_i\rangle_0.
\eeq
We conclude that the leading correction to the vacuum state is
\beq
\vac_1=-\lambda \sum_{n=0}^N 2^{2n}\beta^{\frac{\alpha}{2}-n}\hbar^{n} c_{2n,\alpha}\int\frac{d^{2n-1}p}{(2\pi)^{2n}}\frac{|p_1,\cdots,p_{2n-1},-\sum_{i=1}^{2n-1}p_i\rangle_0}{n\hbar \left(\omega_{\sum_ip_i}+\sum_{i=1}^{2n-1}\omega_{p_i}\right)}.
\eeq

Again, we would like to check that this correction converges at large $N$.  In passing from $N-1$ to $N$,
the change in $\vac_1$ is
\beq
\Delta\vac_1=-\lambda 2^{2N} 2^{2N} \beta^{\frac{\alpha}{2}-N}\hbar^N c_{2N,\alpha}\int\frac{d^{2N-1}p}{(2\pi)^{2N}}\frac{|p_1,\cdots,p_{2N-1},-\sum_{i=1}^{2N-1}p_i\rangle_0}{N\hbar \left(\omega_{\sum_ip_i}+\sum_{i=1}^{2N-1}\omega_{p_i}\right)}
\eeq
which has norm
\beq
|\Delta\vac_1|^2=\lambda^2 \frac{2^{2N}}{N^2}\beta^{{\alpha}-2N}\hbar^{2N-2} c^2_{2N,\alpha}\int\frac{d^{2N-1}p}{(2\pi)^{2N}}\frac{1}{\left(\omega_{\sum_ip_i}+\sum_{i=1}^{2N-1}\omega_{p_i}\right)\omega_{\sum p_i}\prod_{i=1}^{2N-1}\omega_{p_i}}.
\eeq
The $p$ integrals are convergent.  If $\beta$ is proportional to $\hbar$ then this is proportional to $\hbar^{\alpha-2}$.  Again the validity of the semiclassical expansion requires $\alpha>2$.  Models of the kind in Ref.~\cite{conf25} roughly correspond to $\alpha=1$, and this large nonperturbative effect is consistent with the scenario described in that paper.

If $\beta<\hbar$ then at large $N$ these corrections diverge as $(\hbar/\beta)^{2N}$.  Therefore we require that our deformation $\beta$ of the Wick contraction satisfies $|\beta|\geq\hbar$.

\section{The $\phi^4$ Double-Well Model} \label{essez}

Let us consider a simple example corresponding to $\sigma=2$, the $\phi^4$ double-well model.  The potential is
\beq
V=\frac{\lambda}{4}\phi^4-\mu^2\phi^2
\eeq
where $\mu$ is not the mass $m$ at the minima and we have dropped the constant term, which can be set to zero.  Note that our convention is to separate the $m^2\phi^2/2$
 term out of the potential, although this will not affect our conclusions.

Expanding the potential in Hermite polynomials one finds
\beq
V=\frac{\lambda\beta^2}{64}H_4\left(\frac{\phi}{\sqrt{\beta}}\right)+\left(\frac{3\lambda\beta^2}{16}-\frac{\mu^2\beta}{4}\right)H_2\left(\frac{\phi}{\sqrt{\beta}}\right)\nonumber
\eeq
where we have again dropped the constant term.

Now a quick calculation shows that the quantum potential is
\bea
:V:_\beta&=&\frac{\lambda\beta^2}{64}:H_4\left(\frac{\phi}{\sqrt{\beta}}\right):_\beta+\left(\frac{3\lambda\beta^2}{16}-\frac{\mu^2\beta}{4}\right):H_2\left(\frac{\phi}{\sqrt{\beta}}\right):_\beta\\
&=&\frac{\lambda\beta^2}{4}:\left(\frac{\phi}{\sqrt{\beta}}\right)^4:+\left(\frac{3\lambda\beta^2}{4}-{\mu^2\beta}{}\right):\left(\frac{\phi}{\sqrt{\beta}}\right)^2:\nonumber\\
&=&\frac{\lambda}{4}:{\phi}^4:+\left(\frac{3\lambda\beta}{4}-{\mu^2}{}\right):{\phi}^2:\nonumber.
\eea
In other words, ignoring the shift in the constant term, in this case our $\beta$ deformation is simply a shift of the mass.  Of course, nothing more dramatic is required as this model already has an analytic potential.

\section{Finite Stokes Scattering}

In this section we turn to the application that motivated our study.

\subsection{The Model}

We are interested in the $\sigma=4$ P\"oschl-Teller model, corresponding to $\alpha=5/2$ and $\lambda<0$. For convenience, let us define the coupling
\beq
g=\frac{27}{8}\frac{\lambda^2\hbar^4}{m^4}
\eeq
so that $g$ has dimensions of the $1/\sqrt{\hbar}$.

This model is described by the Hamiltonian
\beq
H=\int dx\ch(x)\hsp 
\ch(x)=\frac{\pi^2+(\nabla \phi)^2}{2}+V[\phi] \label{hc}
\eeq
where the potential is, at $|\phi|\leq 1/g$, given by~\cite{sal12}
\beq
V(\phi(x))=\frac{9m^2}{128\hbar^2g^2}\left[1-2\cos \left( \frac{2}{3}{\rm{arcsin}}\left(g\phi(x)\right)\right)  
\right]^4.\label{Vphi}
\eeq
This model exhibits a kink solution
\beq
f(x)=\frac{1}{2g}\left(2+\sech^2\x4 \right)\tanh\x4. \label{fs}
\eeq

Perturbations about the kink are described by the Ansatz
\beq
\phi(x,t)=f(x)+\g(x)e^{-i\omega t}
\eeq
which at linear order in $\g$ reduces to the P\"oschl-Teller equation
\beq
(-\partial_x^2+U(x))\g(x)=\omega^2\g(x) \hsp U(x)=-\frac{5m^2}{4\hbar^2 }\sech^2\x4+\frac{m^2}{\hbar^2}. \label{pt}
\eeq
The solutions are the normal modes of the kink.  These consist of a translation zero mode, three bound shape modes $\g_{S_i}(x)$ and also continuum modes $\g_k(x)$.  We will focus on the highest frequency shape mode
\beq
\g_{S_3}(x)=\frac{\sqrt{5m}}{8\sqrt{\hbar}}\sech^4\x4\left[\sinh\left(\frac{3mx}{4\hbar}\right)-6 \sinh\x4\right]
\eeq
and also the continuum modes
\bea
 \g_k(x)&=&\frac{e^{ikx}}{D_4(k)}\bigg[105\frac{m^4}{\hbar^4}\sech^4\x4+60\frac{m^2}{\hbar^2} \sech^2\x4 \left(12 k^2-2\frac{m^2}{\hbar^2}+7 i k \frac{m}{\hbar} \tanh\x4\right)\nonumber\\
&&+8\left(32k^4-70 k^2 \frac{m^2}{\hbar^2}+3 \frac{m^4}{\hbar^4}+5ik\frac{m}{\hbar}\left(16 k^2-5 \frac{m^2}{\hbar^2}\right) \tanh\x4\right)\bigg]\nonumber\\
 D_4(k)&=&8\sqrt{\hbar}\left(32k^4-70 k^2 \frac{m^2}{\hbar^2}+3 \frac{m^4}{\hbar^4}+5ik\frac{m}{\hbar}\left(16 k^2-5 \frac{m^2}{\hbar^2}\right)\right).
\eea

\subsection{The Problem}

Stokes scattering is the process
\beq
kink+meson\rightarrow kink{}^*+meson
\eeq
in which a kink in its ground state is struck by a perturbative meson which excites one of the kink's internal shape modes.  In the present case, the kink has three distinct shape modes and so there are three distinct Stokes scattering processes in which a single shape mode is excited.

In Ref.~\cite{mestokes}, the Stokes scattering amplitude for exciting the $i$th mode was found to proportional to $
V_{k_1k_2S_i}$ where the three-point vertex factor is defined to be
\beq
V_{k_1k_2S_i}=\int dx V^{(3)}(f(x)) \g_{k_1}(x)\g_{k_2}(x)\g_{S_i}(x) \label{vkks}
\eeq
and $
V^{(3)}(f(x))$ is the third derivative of the potential $V[\phi]$ with respect to $\phi$ evaluated at $\phi(x)=f(x)$.  

In the case of the level $\sigma$ P\"oschl-Teller model, the leading behavior of the potential is a mass term, followed by an interaction \cite{sg2} of the form $|\phi^{2+2/\sigma}|$.  As a result, when $\sigma>2$ the vertex factor $V^{(3)}(f(x))$ diverges when $f(x)$ approaches a minimum of the potential, and so one expects the amplitudes for three-point interactions to diverge.  For example, in the case of the $\sigma=4$ model 
\beq
V^{(3)}(f(x))=\frac{2}{3}\sqrt{\frac{2}{3}}\frac{m^2g}{\hbar^2}\ \sinh\x2 \label{v3eq}
\eeq
which diverges far from the kink, where $|x|\gg \hbar/m$.

In the case of the first three P\"oschl-Teller models, the vertex factors (\ref{vkks}) were shown to be finite in Refs.~\cite{mestokes, hengyuanstokes}.  A similar calculation shows that in the $\sigma=4$ model of interest here, the amplitude for exciting the first two shape modes is finite.  The reason that $V_{kkS}$ is finite despite the divergent $V^{(3)}$ is that the shape modes $\g_{S_i}(x)$ are localized near the kink, and they decrease at large $|x|$ at last as quickly as $V^{(3)}$ increases.

However, one can see from the above formulas that in the case of the third shape mode $S_3$, which has the highest energy and so is the most loosely bound, the $e^{-m|x|/4\hbar}$ decay of $\g_{S_3}$ is not sufficient to compensate for the $e^{m|x|/2\hbar}$ increase in $V^{(3)}$ and so $V_{k_1k_2S_3}$ diverges, leading to a divergent Stokes scattering amplitude.  Critically, this calculation uses an initial quantum Hamiltonian defined using the undeformed normal ordering prescription.

\subsection{The Solution}

To move beyond the vacuum sector, we would like to apply the displacement operator.  Once the expectation value of $\phi$ is much greater than $\sqrt{\beta}$, the Hermite expansion above no longer converges.  But one may Hermite-expand $|\phi^\alpha|$ about another value, $\phi(x)=f(x)$ for some $x$.  Then, applying $::_\beta$ again, one may calculate the various interactions at $x$.

If $\beta\sim\hbar$, this only seems to smooth the potential when $\phi\sim \sqrt\hbar$, and so at distances from the kink such that
\beq
\hbar^{1/2}\sim f(x)-f(\infty)= -\frac{6 e^{-mx/\hbar}}{g}+O(e^{-3mx/2\hbar}) \label{eqeq}
\eeq
which implies
\beq
x\sim -\frac{\hbar \ln{(g\sqrt{\hbar})}}{m}.
\eeq
If one gets closer to the kink, corresponding to a smaller $|x|$, then the unmodified large Stokes rate may be expected.  

Recall that at leading order, up to constants of proportionality, at large $|x|$
\beq
V^{(3)}(f(x))\propto e^{m|x|/2\hbar}\hsp
g_{k}\sim 1\hsp g_{S_3}\sim e^{-m|x|/4\hbar}
\eeq
so that the matrix element is of order
\beq
V_{k_1k_2S_3}=\int dx V^{(3)}(f(x)) \g_{k_1}(x) \g_{k_2}(x)\g_{S_3}(x)\sim \int dx e^{m|x|/4\hbar}.
\eeq
Now this integral is cut off when Eq.~(\ref{eqeq}) holds, yielding
\beq
e^{m|x|/\hbar}\sim\frac{1}{g\sqrt{\hbar}}
\eeq
so that the integrand and therefore the integral itself are of order $O(g^{-1/4})$. 
However, recall from Eq.~(\ref{v3eq}) that the amplitude is proportional to $g$  because cubic terms in $\phi$ are multiplied by a single power of $g$.  As a result, the Stokes scattering amplitude is of order $O(g^{3/4})$, depending on the incoming meson momentum as well as $\beta$.

A similar argument can be applied to meson multiplication, in which a meson strikes a kink and two are emitted.  The interaction factor here is
\beq
V_{k_1k_2k_3}=\int dx V^{(3)}(f(x)) \g_{k_1}(x) \g_{k_2}(x)\g_{k_3}(x)\sim \int dx e^{m|x|/2}\sim g^{-1/2}
\eeq
and so $V_{k_1k_2k_3}\sim O(g^{1/2})$, which also converges at weak coupling.

Recall that in the lower P\"oschl-Teller models with $\sigma<4$, and also in the case of the $\sigma=4$ model with the shape modes $S_1$ and $S_2$, the amplitude for Stokes scattering is of order $O(\lambda)$.  Now we have learned that in the $\sigma=4$ model, the least bound mode $S_3$ extends so far into the bulk that it sees the nonanalytic $|\phi|^{5/2}$ potential.  When the resulting divergence is cut off by $\beta$, nonetheless a $\beta$-dependent contribution from the divergence remains.  This divergent contribution to Stokes scattering leads to an amplitude that is not of order $O(g)$ but instead of order $O(g^{3/4})$ when $\beta\sim O(\hbar)$.

In general we expect that the more loosely bound a mode is, the further it extends into the bulk and so the more it is affected by the bulk's divergent albeit cutoff nonlinearities.  As a result, the most loosely bound modes enjoy Stokes scattering amplitude at lower orders than $O(g)$.  

\section{Comments}

It has long been known \cite{gj} that, in the case of (1+1)-dimensional models with polynomial potentials, normal ordering is sufficient to remove all ultraviolet divergences.  However, as described in Ref.~\cite{colemansg}, normal ordering itself is not unique, but rather depends on a mass scale.  As this mass scale is varied, the normal-ordered terms themselves vary.  

In the present paper, we have used Coleman's deformation to cure a divergence that appears in the derivatives of the potential already classically.  We found that even a finite, but sufficiently large, modification of the mass is sufficient to remove divergences in the case of a noninteger power law model. 

This deformation was required to obtain a finite amplitude for Stokes scattering in the $\sigma=4$ model, where the falloff of the least bound mode far from the kink was slower than the increase in the third derivative of the potential.  In other words, the least bound shape mode did not stay close enough to the kink to be protected from the divergence in the vacuum.  However, the deformation was in fact already required to obtain a finite amplitude for meson multiplication in the $\sigma=3$ model, as the meson three-point interaction is not confined to be near the kink.  As a result, with this deformation in hand, both meson multiplication and Stokes scattering amplitudes may be calculated for any P\"oschl-Teller model kink, or more generally kinks in other models with such noninteger power law potentials, such as the Rosen-Morse potentials \cite{rm1,yuanrm}.

In recent years, numerous examples \cite{qm0,qm1,qm2,qm3,qm4} have emerged of small quantum effects which have phenomenologically important effects on solitons.  The deformation considered in this paper affects the space far from the soliton, but in doing so it protects the soliton's internal modes from being excited by distant radiation.

\appendix
\section{Hermite Expansion} \label{happ}

Using the completeness of the Hermite polynomials
\beq
\int_{-\infty}^\infty dx H_{2n}(x)H_{2m}(x)e^{-x^2}=\sqrt{\pi}2^{2n}(2n)!\ \delta_{mn}
\eeq
to normalize the coefficients, one finds
\beq
c_{2n,\alpha}=\frac{\int_{-\infty}^\infty dx H_{2n}(x) |x^\alpha| e^{-x^2}}{\sqrt{\pi}2^{2n}(2n)!}. \label{cpad}
\eeq
To find the numerator, on first uses the substitution $y=x^2$ to evaluate
\beq 
\int_{-\infty}^\infty dx x^{2m}|x^\alpha|e^{-x^2}=
2\int_{0}^\infty \frac{dy}{2\sqrt{y}}y^{m}y^{\alpha/2}e^{-y}=\Gamma\left(m+\frac{\alpha+1}{2}\right).
\eeq
The identity
\beq
H_{2n}(x)=e^{x^2}\frac{\partial^{2n}}{\partial x^{2n}}e^{-x^2}
\eeq
then leads to
\bea
\int_{-\infty}^\infty dx H_{2n}(x) |x^\alpha| e^{-x^2}&=&\int_{-\infty}^\infty dx \left( \frac{\partial^{2n}}{\partial x^{2n}}e^{-x^2}\right) |x^\alpha| =\int_{-\infty}^\infty dx e^{-x^2} \frac{\partial^{2n}}{\partial x^{2n}}|x^\alpha|\\
&=&(\alpha)_{2n}\int_{0}^\infty dx e^{-x^2} |x^{\alpha}|x^{-2n}=(\alpha)_{2n}\Gamma\left(-n+\frac{\alpha+1}{2}\right).\nonumber
\eea
Substituting this into (\ref{cpad}) we find
\beq
c_{2n,\alpha}=\frac{2^{-2n}(\alpha)_{2n}}{\sqrt{\pi}(2n)!}\Gamma\left(-n+\frac{\alpha+1}{2}\right)=\frac{2^{-2n}(\alpha)_{2n}}{\sqrt{\pi}(2n)!}\frac{\Gamma\left(\frac{\alpha+1}{2}\right)}{\left(\frac{\alpha-1}{2}\right)_n}=\frac{1}{\sqrt{\pi}(2n)!}\left(\frac{\alpha}{2}\right)_n\Gamma\left(\frac{\alpha+1}{2}\right).
\eeq

\section* {Acknowledgement}

\noindent

This work was supported by the Higher Education and Science Committee of the Republic of Armenia (Research Project No. 24RL-1C047).
HY Guo was supported by Sun Yat-sen university international Postdoctoral Exchange Program and also supported by Research Projects Developed by the Lanzhou Theoretical Physics Center/Gansu Provincial Key Laboratory of Theoretical Physics (Research Project: NSFC Grant No. 12247101).  S.B. and H.G. are supported by the INFN special research project	grant ``GAST'' (Gauge and String Theories).

\end{document}